\documentclass{JHEP3}
\usepackage{epsf}
\usepackage{amsmath}
\usepackage{amsfonts}
\usepackage{amssymb}

\newcommand{\pa}{\partial}
\newcommand{\tr}{{\rm tr}}
\newcommand{\comment}[1]{}

\newcommand{\pasl}{\pa\kern-.55em /}

\newcommand{\ksl}{k\kern-.55em /}

\DeclareFixedFont{\xiiss}{OT1}{cmss}{m}{n}{12}
\DeclareFixedFont{\ixss}{OT1}{cmss}{m}{n}{9}
\DeclareFixedFont{\cmrnine}{OT1}{cmr}{m}{n}{9}

\newcommand{\CCs}{\hbox{\ixss C\kern-.4emI}}
\newcommand{\ZZs}{\hbox{\ixss Z\kern-.4emZ}}

\title{Shape and holography: studies of dual operators to 
giant gravitons.}
\author{David Berenstein \thanks{dberens@ias.edu}\\
 School of Natural Sciences, 
Institute for Advanced Study, 
Princeton, NJ 08540, USA
}

\abstract{In this paper we study the conjectured
dual operators to a 
near maximal  
giant graviton and their open string fluctuations in the large $N$ 
limit. 
Using matrix model estimates we 
show that the spectrum of states near the D-brane operator
is consistent with a Fock 
space of open plus closed string states. 
We also give an argument that 
these operators, in spite of having large R charge of order $N$,  
are amenable to being studied with standard
perturbative techniques, which organize themselves in a 
$1/N$ expansion. Also the
spectrum of operators dual to 
massless fluctuations on the D-brane is shown to be
protected
from weak to strong coupling at leading order, so it is possible to 
read the shape of the dual operator by understanding how the 
spherical harmonics of the D-brane fluctuations appear.}

\begin{document}

\section{Introduction}

The AdS/CFT correspondence \cite{Malda} is a remarkable duality between a quantum 
theory of gravity (a string theory or M-theory) 
on a particular background 
and an ordinary quantum conformal field theory in fewer 
dimensions. 
The most studied example involves type IIB string theory on 
$AdS_5\times S^5$ with $RR$ flux 
and ${\cal N}=4$ SYM in four dimensions on the 
Lorentzian manifold $S^3\times R$. 

These are examples of holography \cite{'tHooft,Susskind}: 
it is believed that any theory of
quantum 
gravity is a non-local theory \footnote{In a theory of quantum gravity 
there are no local observables
because  quantum 
fluctuations destroy any notion of position at arbitrarily small 
scales.}, and that the degrees of freedom of a region of space time are
captured completely by the boundary of the region. 
If this is true, then one has to understand how some notion of
locality can appear so that the theory has some 
classical limit as a theory of gravity on a large space time.

To understand the concept of locality one needs a laboratory 
for holography where these notions can be made sense of. In this case 
the ${\cal N}=4$ SYM theory dual to IIB strings on $AdS_5\times S^5$
provides the natural setup to ask these questions.

To have an approximate 
notion of a point in a theory with quantum gravity, 
we need a probe that can 
be associated with such a point. To make the notion of a 
point concrete 
at some scale (lets say $\Delta$), 
one needs an object whose fluctuations in position are
much shorter than the given scale, and which deform the background metric
very little. This gives us a bound $M> \Delta^{-1}$. Also, the 
gravitational radius of the probe should be smaller than $\Delta$. 
From here another inequality follows, where $M<f(\Delta, M_{pl})$
which depends on the dimension of space time.

Putting those bounds together tells us that $\Delta > 1/M_{pl}$
In general this is the best we can hope for. There should be no 
notion of approximate classical geometry at the Planck scale. 
In general, to define approximate locality we should be able to 
take a classical limit in the following sense: consider a family of 
metrics for space time which vary only in their scale $R$
(the radius in Planck units). 
The classical geometry will become a better approximation to the 
quantum system when the radius of the geometry goes to infinity.
In this family one would want a probe which is heavy (lets say of 
the order of the Planck scale, or whose scale is fixed as we take $R$ 
to infinity). The position of the 
heavy probe will give us a notion of classical
point in the limit $R\to \infty$, because the fluctuations of the 
object will become suppressed with respect to the scale $R$.

If we want to understand locality in a (weakly coupled) 
string theory, before we reach 
the Planck scale we get to the string scale, and here classical 
geometry is replaced by a more quantum version of it: stringy 
geometry (two dimensional CFT's), but this is still far from 
giving geometrical concepts at the Planck scale.

Therefore, if we 
want to understand geometry at the string scale we need a probe that is 
heavy with respect to the string scale, but which is still tractable.
The natural object with these properties is a D-brane. These are 
non-perturbative objects of string theory, and therefore are 
heavy. 
In the weak coupling limit one should consider them as boundary 
states of the world sheet CFT, and in the large radius limit they
become geometric objects (the locus where strings end).

These are the objects one would like to study to define 
locality at the string scale. However, not all D-branes are the same. 
Some are unstable, some are extended, some configurations are 
anomalous (their D-brane charge has nowhere to go).

From the space time point of view, all of this seems standard. However, 
when we mix the ingredient of holography and ask the same questions 
in the holographic dual of a gravitational 
theory, we find that we need all of the above 
considerations before we can make sense of what it means for a brane 
to have some position (or if the brane is extended some shape).
 Even after we decide how to understand these 
questions we need to be able to trust the calculation 
on the holographic dual of the theory and match the candidate 
quantum state on the boundary with a semi-classical object in the 
geometry. 

The SYM theory is tractable only at weak coupling, 
via perturbation theory.
The string theory is tractable if the 
background is weakly curved in string units, and if the string
scale 
is smaller than the gravitational scale. This regime requires that the
dilaton be small everywhere, and that the geometry is very large.

Both statements of tractability are determined by the 
parameters that define the theory. 
The parameters of the SYM theory are the gauge group $U(N)$, and the 
complex gauge coupling $\tau\sim \frac 1 {g^2_{YM}}+i\theta$.
On the string theory side the parameters are the closed string coupling 
$\tau\sim \frac 1 {g_s} + i a$
 and the radius of the $AdS$ space in string units, $R$.
The transformations under $S$-duality identify the two complex 
couplings with each other.
 
Quantization of the RR flux on the AdS space
produces a number for the radius of 
the sphere in string units, which is of the order 
$R^4\sim g_sN \sim g_{YM}^2 N$. This is the 't Hooft coupling of the 
gauge theory, and it dominates the perturbative expansion
of the theory in the large $N$ limit \cite{'tHooftplan}.
 If we want quantities to be calculable 
using the superstring theory we need to take a limit where the 
string theory is weakly coupled $g_s \to 0$ or very small, and where 
the radius $R$ is very large in string units. This implies that 
we should take $N$ very large. Moreover, the strict limit 
$R\to \infty$, $g_s\to 0$ is the supergravity limit of 
string theory. This regime of calculability is the strong 
't Hooft coupling.

Similarly, for calculability in the SYM theory, we would prefer to 
have a situation where perturbation theory can be trusted. The 
perturbation theory of the field theory is controlled by the 't Hooft 
coupling $g_{YM}^2 N$, so the perturbation expansion can be trusted 
if $g_{YM}^2N$ is small.

The AdS/CFT duality is therefore a 
strong-weak coupling duality, in the sense that calculability
forces us to different (in principle incompatible) 
corners of the parameters.
 For many 
processes it is impossible to extrapolate the dependence on the 
coupling and we have to trust the duality to do a calculation.

The aspect of this setup that makes checks of duality possible is
that the theory is supersymmetric, and there are
protected objects (BPS states) for which it is possible to 
compare results at strong and weak coupling.
In general we will define {\em almost BPS states} as those for
which it is possible to extrapolate between weak and strong 
coupling.

If one wants to study locality with D-brane probes, then one needs 
to be able to control the calculation, and one wants to have BPS 
protected objects.
There are again 
non-perturbative states on the $AdS\times S^5$ spacetime which are 
protected by supersymmetry and correspond to D-brane states. These 
have been labeled giant gravitons \cite{giant}, and the simplest ones
are given by a $D3$ brane which wraps a round $S^3$ in $S^5$ and 
is spinning on the $S^5$. Other configurations have been 
considered \cite{Hetal,GMT}

These are not point-like however. They are extended in the the $S^5$ 
direction, and semi-classically they have a shape. What is 
the description for these D-branes in the holographic dual?
How can we test their shape? 

The key to answering these questions is on to how to 
compare calculations on both sides of the duality: the AdS/CFT 
dictionary.
If we say that two quantum theories are equivalent then we should  
be able to match the states and operators (correlation functions) 
between both theories. We should be able also to extrapolate these 
results between weak and strong coupling.

For $AdS_5\times S^5$, which has 32 supersymmetries,  
any massless state 
in ten dimensions is BPS (they preserve half 
of the supersymmetries), and it has been possible to match the 
spectrum of gravitational fluctuations with the spectrum 
of totally symmetric traceless single 
trace operators 
$\tr(\phi^{i_1}\dots \phi^{i_k})$ and their descendants under the 
full supersymmetry algebra \cite{Wittenhol,GKPads}. Their dimension can be 
calculated in the free field theory, so one can make a complete match
between gravitational states on both theories. 
Also, one can test the 
interactions, and there are some protected correlation 
functions of these operators which are calculable 
and non-renormalized \cite{seietal}.

Matching the supergravity states alone does not 
imply that one has a full string theory.
The large $N$ limit of 't Hooft for the quantum field theory
does suggest a string theory, but it is not tractable for most 
applications.  Miraculously there is 
a way to find some tractable strings on the $AdS$ 
geometry. It was understood in \cite{BMN} that one could take a 
geometric plane wave limit in the holographic 
dual theory. It turned out that 
for some special states 
the perturbation theory in the 't Hooft coupling is
parametrically suppressed by the quantum numbers of the state.
This parametric suppression made it possible to extrapolate the 
perturbation theory from weak to strong coupling, and the suppression 
was tied to the state being almost supersymmetric (almost BPS states).

One can define a parameter which measures the deviation from being
 BPS for a given state.
\begin{equation}
\eta = \frac{|\Delta -J|}{J}
\end{equation}
where $J$ is one of $R$-charges of the state.
This parameter $\eta$ 
does not necessarily control the perturbation theory, 
but $\eta=0$ is a BPS state. A state is usually almost
BPS if in the geometric limit $\eta$ goes to zero. 
In the plane wave limit $\eta \sim N^{-1/2}\to 0$, and the 
BPS states can be matched with graviton multiplets on the plane 
wave.

Similarly if we consider a BPS brane with some stringy 
excitations on it, one expects that in the geometric limit the 
D-brane dominates the mass, charge and back reaction of the geometry, 
so $\eta\to 0$. This is even true if we consider also some closed 
string states in the mix.

The limit therefore is a standard image of D-branes: in the 
spectrum of states 
there can be open strings and closed strings. The space of states should 
be approximated  in the perturbative string theory limit by a Fock space 
of open and closed string states.

Now, because the D-brane is an extended object, in the large radius 
limit one can consider also small fluctuations of shape for the 
D-brane. These fluctuations are geometrical, and their spectrum becomes 
independent of the radius in units of $R$. (The tension of the 
D-brane will appear suppressing the interactions between 
these fluctuations.) 
These fluctuations  depend on the shape 
of the D-brane. To see a shape in the holographic dual
we should be able to find not
just the D-brane state alone, but also the fluctuations around it.
Examples of this notion of shape 
can be found in various different situations, although in most cases
the dual holographic theories are not understood very well.
To be more specific, in Matrix theory one could define 
spherical membranes by certain configurations of matrices 
\cite{KT,Sjr}. The spherical harmonics could be reproduced by 
understanding the perturbative spectrum of states on the membrane
configuration. However, the spherical membrane itself is unstable 
for flat space. This can be remedied for wrapped membranes on some 
2-sphere \cite{BC}, and one recovered the right 
spectrum of states in a large $N$ limit. This effect 
can also be remedied by 
introducing a flux which stabilizes the spherical membranes. There 
is a matrix model for M-theory on a maximally supersymmetric
plane wave which has these 
properties \cite{BMN}. These spherical membranes are remnants of
giant gravitons of 
$AdS_{4,7}\times S_{7,4}$.
One can  also find evidence for the existence of 
five-branes \cite{MV} by understanding that some fluctuations on the brane 
world-volume are protected by supersymmetry
\cite{KiP,DSV}, and therefore their spectrum
can be calculated even when the matrix model is strongly coupled.
Another example where one can ask what the shape of branes is, 
arises from exploring boundary states for WZW CFT's in the large
level limit \cite{MMS}. There one does an overlap with a localized 
wave function to find the location of the branes. However, this
example is strictly understood only on the target space and not 
on it's holographic dual theory: it is an example of how stringy 
geometry becomes classical in the large radius limit.
 Finally, 
the string quantization in 
the plane wave limit \cite{BMN} can also be considered as a semi-classical 
expansion of an extended object (the fundamental string)\cite{GKP}. One
can think 
of this expansion in terms of a purported T-duality along a circle 
of $S^5$. This duality 
exchanges momentum with winding, and a string with a lot 
of momentum on the $S^5$ could be thought of as a long string (with
high winding) on the T-dual circle. This would justify a geometric 
interpretation of the string as an extended object 
in the T-dual geometry.

For the giant gravitons in $AdS\times S$, 
the spectrum of fluctuations has been calculated in \cite{Jetal}. 
Surprisingly, the quantization of modes does not depend on how
big the giant graviton is, and it is compatible with 
dimensions that could be obtained from free field theory. This
fact suggests that these states might be protected \cite{BHK} in 
the geometric limit.

Dual operators to giant gravitons on $AdS_5\times S^5$
have been conjectured in the 
literature \cite{BBNS}.
Operators dual to giant gravitons with excitations (in the plane 
wave limit) have been 
conjectured in \cite{BHLN} based on the description of states in 
\cite{BHK}.
These dual operators 
are determinant and sub-determinant operators, whose 
dimension is large (of order  $N$). For these states, non-planar 
diagrams dominate over planar diagrams, so expressions with traces 
mix too much to be useful.
The tests that have been performed on these states are mostly of 
group theoretic nature, in that it has been shown 
that they have the correct $R$ charge and factorize better than traces
of the same dimension \cite{CJR}. Also, it has been argued that BPS
correlation functions with 
gravitons and a giant graviton produces expansions in topologies with 
boundaries\cite{AABF}. These tests are of BPS correlators, and depend just
 on the free field structure of the theory. 
However, making a systematic expansion with the methods 
available in the literature seems hard to do.

Of the tests above, only the last one starts to look like a 
test that would show that the conjectured dual giant gravitons 
truly represent D-branes. 
However, since the calculations are BPS protected,
 it is hard to understand
what is the nature of the results that one is obtaining, and one 
also has to worry about mixing of states.

 The objective of this paper is to make the claim that these operators 
are giant gravitons as solid
as possible: 
 to show that the notion of shape and open string excitations
emerges asymptotically in the 
large $N$ limit. We will also show that the 
operators in question, 
describing excited D-branes with some spherical harmonics,
 have 
protected dimension in the strict large $N$ limit, at small but finite
$g_{s}$ and in the leading approximation. We will also show that the 
physics of these objects can be organized in a $1/N$ expansion.
To simplify the combinatorial problem, 
we will limit ourselves to configurations of a
single giant graviton with 
(near) maximal angular momentum on the $S^5$. 

The technical aspects of this paper are mostly combinatoric: how to 
organize free field diagrams and the perturbation theory around 
the D-brane states. 

The conjectured dual operators are very similar to baryonic operators.
The main  difficulty in dealing with  baryonic 
objects in the large $N$ limit is that they are usually not amenable to 
a standard (diagrammatic)
perturbation theory \cite{Wittenbar}, although there are other 
approximations which can be useful. 
Perturbative corrections tend to pile up with 
powers of $N$, just from combinatorial multiplicity of the 'partons'.
 Since the giant gravitons are BPS, there is a 
possibility that most of these diagrams actually
cancel, and that all
corrections are due just to the deviation from the object 
being BPS. 
We will show that this second intuition 
turns out to be correct, and that one 
can organize the perturbative diagrams systematically for these baryonic 
like-objects.
This procedure involves turning the problems above into 
standard Gaussian matrix model correlators in the  $1/N$ expansion, 
where the matrix model correlators do not scale with $N$.

Since giant gravitons are similar to (di)-baryons, the 
combinatorial structure described above can be generalized to 
many other CFT's (see for example 
\cite{Wittenbads,GRW,GK,BHK,Beasley,Ouyang,IW,HM}) 
and similar results should hold there even if they 
are not ultimately calculable. 

The paper is organized as follows:

In section \ref{sec:Fock} we give a combinatorial construction of a
Fock space of open-string states on the D-brane operator. 
In particular we detail how the structure of spherical harmonics on the 
D-brane is produced. We show 
also that some combinatorial boundary conditions appear for the string 
spectrum, which begins to suggest the notion of a D-brane.
The techniques to show the structure 
are based on turning the problem into a standard
Gaussian 
matrix model correlation function, where the number of letters in a
word does not scale with $N$ in the large $N$ limit.

Next, in section \ref{sec:oneloop} we calculate the leading order 
anomalous dimension of a giant graviton with massless excitations. We 
show that the anomalous dimension is suppressed in the large $N$ limit, 
even in the limit of finite $g_{YM}$.

We then consider in \ref{sec:org} the problem of organizing the
$1/N$ expansion for all of these states. We find a heuristic argument 
that suggests that there is a well defined $1/N$ expansion. 
The argument is not complete and requires checking the detailed 
structure of the states at higher loop orders.

We end the paper with some conclusions and an appendix on the 
combinatoric identities that are used throughout the paper and make the 
problem tractable.

\section{Fock space of fluctuations on a single giant operator}
\label{sec:Fock}

A standard picture of a D-brane is as an object in a string background 
where strings can end. The D-brane is a non-perturbative object,
but there is a well defined perturbation theory around the object 
itself.
Perturbatively in string theory, 
in the presence of a 
D-brane, the Hilbert space of states is a Fock space of closed string 
plus open string states. This structure gets corrected when 
interactions are turned on: for example, open and closed string 
states mix, there can be a (non-perturbative) 
stringy exclusion principle, etc.

In order to make these statements one has to consider 
the fact that they are only an approximation in the geometric 
limit (this includes stringy geometry limits). 
Thus all of these statements should be corrected by the 
genus expansion, which in the $AdS_5\times S^5$ geometry, is the 
$1/N$ expansion in the dual field theory.

Of the spectrum of strings on a D-brane, some are 
massless modes on a D-brane (this statement makes sense in the 
geometric limit, where
the mass scale of these states is much smaller than the string 
scale). 
The
effective action for these modes is captured by the Dirac Born 
Infeld action. A calculation of the spectrum of these modes 
for the giant gravitons we will study has been
performed in \cite{Jetal}.

It turns out that the spectrum of fluctuations (in the geometric limit) 
is independent of the size of the giant graviton, and some of 
these states saturate a BPS bound for some of the R-charges. 
Since the D3-branes do not couple directly to the dilaton, varying the
coupling should not affect it's calculation, so it is conceivable that so 
long as gravity is weakly coupled with respect to the ADS scale 
($N$ is large), the calculation should be reliable.
This suggests that 
these states do not change their spectrum 
when we take the zero coupling limit, 
so it should be possible to describe them simply in the 
free field theory (large $N$) limit.

These states, if we understand them in the dual theory,
give rise to the structure of spherical 
harmonics on 
the giant graviton. Given this knowledge, it is possible to begin
building the full set of string states ending on the 
D-brane.

We will now check that one can write a Fock space of excitations 
of a giant graviton 
operator which is consistent (on group theoretical grounds) with 
the Fock space of fluctuations of a membrane with the shape of 
$S^3$.

To begin, 
the BPS spherical membrane is conjectured to be given by sub-determinant 
operators
\begin{equation}
{\cal O}^{N-k}_{D3}  = \epsilon^{\mu_1, \dots \mu_{N},  } 
\epsilon_{\rho_1, \dots \rho_{N}} 
\phi_{\mu_1}^{\rho_1}\dots  \phi_{\mu_{N-k}}^{\rho_{N-k}}
\delta_{\mu_{N-k+1}}^{\rho_{N-k+1}}\dots \delta_{\mu_{N}}
^{\rho_N}
\end{equation}

We will use the following abbreviated notation
\begin{equation}
{\cal O}^{N-k}_{D_3} = 
\epsilon\epsilon(\phi, \dots,\phi, \overbrace{1, \dots ,1}^k)
\end{equation}
to denote the operator. 
The $N-k$ subscript counts the number of indices saturated by the field
$\phi$. In general we will treat the $\epsilon\epsilon$ convention as a 
multilinear operator with $N$ slots, where we can insert various fields.
The purpose of this notation is mainly to
keep track of how the indices are contracted without writing the 
indices explicitly. When the need arises we will restore the 
indices. This is also explained in the appendix.

The normalization of the operator ${\cal O}^{N-k}$ can be calculated 
with the combinatorial identities in the appendix to 
give
\begin{eqnarray}
<O(x)\dagger O(y)>|x-y|^{2(N-k)} &=&  (N-k)! 
\left[ \epsilon^{\mu_1,\dots,\mu_{N-k},\rho_1,\dots,\rho_k}
\epsilon_{\mu_1,\dots,\mu_{N-k},\sigma_1,\dots,\sigma_k}\right.\\
&& \left. \epsilon^{\nu_1,\dots,\nu_{N-k},\sigma_1,\dots,\sigma_k}
\epsilon_{\nu_1,\dots,\nu_{N-k},\rho_1,\dots \rho_k}\right]\\
&=&  (N-k)!^3 \delta^{[\rho_1,\dots,\rho_k]}
_{[\sigma_1,\dots,\sigma_k]}\delta^{[\sigma_1,\dots,\sigma_k]}
_{[\rho_1,\dots,\rho_k]}\\
&=&(N-k)!^3 k! 
\delta^{[\rho_1,\dots,\rho_k]}_{[\rho_1,\dots,\rho_k]}\\
&=& (N-k)!^3 k! {N \choose k} k!\\
&=& (N-k)! [(N-k)!k! N!]
\end{eqnarray}
We will also use the following simplifying notation
\begin{equation}
\epsilon^{\mu_1,\dots,\mu_{N-k},\rho_1,\dots,\rho_k}
\epsilon_{\mu_1,\dots,\mu_{N-k},\sigma_1,\dots,\sigma_k}
\equiv \epsilon\epsilon_{k,k}()
\equiv (N-k)! \delta_{k,k}()
\end{equation}
where $k$ indicates the number of indices that are left 
uncontracted on each epsilon symbol, and $\delta()$ 
is completely antisymmetric in it's upper and lower indices
\footnote{
Hopefully it will be clear which convention is being used if one 
compares this notation with the Kronecker 
$\delta^\mu_\nu\sim \delta_{1,1}$ 
which has just one upper and one lower index.}. 

The contractions above can then be rewritten as
\begin{equation}
\epsilon\epsilon_{k;k}\epsilon\epsilon_{k;k}
= (N-k)! N! k!
= (N-k)!^2 \delta_{k;k}\delta_{k;k}
\end{equation}
in the notation above, contraction of indices between 
objects is
performed by using the same letter $k$. If we want to keep the indices 
separate then we will use a notation where we consider changing the
subscripts to $\bar k$, etc.

The expression 
$\delta_{k;k}\delta_{k;k}$ has a very nice $1/N$ 
expansion if $k$ is fixed and does not scale with $N$, namely
\begin{eqnarray}
\delta_{k;k}\delta_{k;k} &=& k!N^k (1-1/N)(1-2/N)\dots (1-(k-1)/N) 
\\ &=&
k! N^k \left[1 - {k\choose 2} N^{-1}
+\left({k\choose 3} +{k\choose {2 ,2}}\right)N^{-2}\right.\\
&&
\left.-\left({k\choose 4} + {k\choose {3, 2}}+{k\choose{2, 2, 2}}\right)
N^{-3}+\dots\right]
\end{eqnarray}
in the expression above we use the following notation for the 
multinomial coefficients
$$
{l \choose n_1, \dots,n_s} = \frac{l!}{n_1!n_2!\dots n_s! (l-\sum n_i)!}
$$
and the power of $N^{-1}$ accompanying ${k \choose n_1,\dots n_s}$
in the expansion is $\sum_i (n_i-1)$.
Keeping $k$ finite and letting $N$ go to infinity puts us very near 
the maximal giant operator. Notice that the (finite) 
series is dominated by the 
first term so long as $k^2<N$, so it is possible to use this expansion
even in this limit. This scaling is reminiscent of the dimension 
of operators in the Penrose limit
of $AdS_5\times S_5$ \cite{BMN,KPSS,BN,seven}.
For the remainder of the paper we will 
consider only this possibility of $k$ small, where the above $1/N$ 
expansion is well behaved.

This expansion comes from noticing that 
\begin{equation}
\delta^{[\rho_1\dots \rho_k]}_{[\mu_1\dots\mu_k]}
= \sum_{\sigma} (-1)^{|\sigma|}
\delta^{\rho_1}_{\mu_{\sigma(1)}}\dots \delta^{\rho_k}_{\mu_{\sigma(k)}}
\end{equation}
is a sum of Kronecker $\delta$ over the permutations of the set 
${1,\dots ,k}$. Each term establishes a different identification 
between 
the upper and lower indices of the $\delta_{k;k}$. 
When we contract two of these tensors we have the composition of two 
of these permutations, with all indices contracted, which is again a 
permutation. Each such 
permutation identifies various indices among themselves. The factor
of $N$ appearing in the expansion is $N^{\#(\sigma)}$, the number of 
cycles in the permutation, as the indices associated to each cycle are 
identified. The highest one has all cycles of order 
one. The next order has one cycle of order $2$ and all others of order 
$1$. At the third level 
we can 
get either two cycles of order $2$ or one cycle of order $3$, 
etc. The factor of $k!$ comes about because we sum over
two sets of permutations, but only the total composition 
of the permutations matters. 

This identity can be generalized to having contractions with arbitrary 
matrices in the following form:
\begin{equation}
\delta_{k;k}\delta_{\tilde k,\tilde k}
(M_1, \dots, M_k ;\tilde M_1, \dots \tilde M_k) 
\end{equation}
where the $k$ upper indices and $\tilde k$ lower indices are contracted 
with the $M$, and the
$k$ lower indices and $\tilde k$ upper indices are contracted with
the $\tilde M$.
This is equal to an alternating sum of all possible combination of
traces
\begin{equation}\label{eq:sum}
\sum_\sigma \prod_ {i} \tr(M_i \bar M_{\sigma(i)}) -
\sum_{u, t \sigma} \tr(M_t \bar M_{\sigma(t)}M_u
\bar M_{\sigma(u)})\prod_{i\neq t,u}\tr(M_i \bar M_{\sigma(i)})
+\dots
\end{equation}
where in each trace we have an equal number of
distinct alternating $M, \bar M$ in all possible 
combinations (modulo cyclicity of the trace).
This is described in detail in the appendix.

The sign is determined by the number of pairs of matrices 
$M_i\bar M_j$ which are fused into larger traces
($k$ minus the total number of traces).
If all the $M, \bar M$ are set to the identity, then we get 
the same sum as described above. The factor of $N^{k}N^{-s}$
comes from having exactly $k-s$ traces of the identity.
The factor of $k!$ is then the number of ways to pair
a set of $k$ upper indices with $k$ lower indices. This is 
the number of times that $\tr(1)^k$ appears in the sum.

For the remainder of the paper we will set $k$ finite. From the 
above expressions it is easy to see that when $k$ is finite the 
total number of traces to consider is finite. 
If for each trace we associate
a factor of $N$, the terms with the most traces dominate, so long 
as the number of terms with fewer traces is suppressed with respect to 
$N$. This again produces the estimate $k^2\leq N$.

Now, we are ready to start writing operators which correspond to 
massless excitations on the
D3-brane worldvolume. For this, let us consider the following operators
\begin{equation}\label{eq:onequanta}
{\cal O}^{N-k}_{a_n^\dagger D3} = 
\epsilon\epsilon(\phi,\dots,\phi,\overbrace{1,\dots,1}^{k-1},Z^n)
\end{equation}
We will claim that this is an operator which corresponds to a D3-brane 
with one quantum of momentum $n$ on the worldsheet of the D3-brane.
Since the field $\phi$ is invariant under an $SO(4)$ subgroup of the 
$SO(6)$ R-symmetry, we can think of this state as a highest weight 
state with respect to the $S0(4)\sim SU(2)\times SU(2)$ algebra 
(which is of spin $(n/2,n/2)$ with respect to $SU(2)\times SU(2)$).
Already this counting shows that there is exactly one state 
for each spherical harmonic of a scalar field in $S^3$ (the spherical 
harmonics come from traceless
tensor products of the $(1/2,1/2)$ representation).

Since this state is a chiral operator in the free field theory,
 it is possible for it to be 
protected. Of the transverse fluctuations of the D3-brane there 
are four polarizations along $AdS$, and two polarizations along
the $S^5$. The states above have no polarization along the AdS
directions because they are not charged under the 
$SO(4)$ rotation symmetry. Thus they should 
correspond to transverse fluctuations along the $S^5$. These combine 
to form a complex scalar field, and the states above correspond 
to one of the polarizations of this scalar field. For each 
momentum there will be two polarizations, but only one is BPS, 
the other one is anti-BPS. Since the above states are chiral, they 
should be matched with the BPS polarization.

Now, we need to remember that the D3-brane is a half-BPS state, so it
should be possible to act on this state with an excitation
with the supersymmetries that 
the D3-brane left unbroken. These massless states on the D-brane 
are acted on by the unbroken supersymmetries and 
generate the full multiplet of a massless field on the $S^3$.
The D3-brane is killed by all of the conformal supersymmetries, 
and half of the standard supersymmetries of the superconformal algebra.
The conformal supersymmetries kill any superconformal primary field, 
so we only need to worry about the standard supersymmetries. There are 
$16$ in total. Half of those are broken by the D-brane, 
which leaves us with $8$
unbroken supersymmetries. Of these, half are unbroken if the state is 
BPS, so it gives rise to 
$2^4$ states in a multiplet modulo descendants.
It is found that there are $2^4$ conformal primaries in each 
such superprimary operator, without taking into account the 
supersymmetries broken by the D-brane istself. 
Half of these are fermionic, and 
half bosonic, and it is the particle content of 
${\cal N}=4$ SYM. 
This is enough to produce the full set of polarizations 
for the supersymmetric multiplet on the D3-brane, so counting
the BPS states plus their descendants with respect to the unbroken 
supersymmetries generates the full spectrum of 
polarizations of the fluctuations of the D-brane and 
can be calculated from the above operator. Given this identification, 
we can rest assured that we have the complete counting of 
polarizations that we need to match the supergravity 
calculation.

Similarly, a state with two BPS quanta will be argued to be given by
\begin{equation}\label{eq:twoquanta}
{\cal O}^{N-k}_{a_m^\dagger a_n^\dagger D3} = 
\epsilon\epsilon(\phi,\dots,\phi,\overbrace{1,\dots,1}^{k-2},Z^m,Z^n)
\end{equation}
and essentially, anytime we want to add one quantum we 
substitute one of the $1$ matrices in the operator by 
the word $Z^{n_i}$ if we want it to have momentum $n_i$.
Here we have specialized to states which are BPS
and have maximum spin $(n/2,n/2)$,
but it is very easy to write other states by using the 
$SO(4)$ group theory. The words that are substituting the $Z^m$ 
with the BPS
transverse scalar polarization along the $S^5$ (so long as they do not contain $\phi,\bar\phi$,
are fully symmetric and 
traceless in the 
$S0(4)$ indices.  

Now, there are a few details which we need to be careful about.
First, we need to show that we have a Fock space of states (particles 
with the same quantum numbers are indistinguishable).
 For 
example $a_n^\dagger a_m^\dagger |D_3> \sim a_m^\dagger a_n^\dagger
|D_3>$. This identity follows because  the $\epsilon$
symbols are totally antisymmetric in their indices, so that the 
$\epsilon\epsilon$ operator is completely symmetric 
in it's entries. In essence, the states have the correct statistics
to be interpreted as a Fock space of bosons. Similarly with fermionic 
operators, there is an extra $-$ sign from reordering 
the operator, which is inherited from the statistics of the 
fermionic fields on the field theory.

Also, there are states with closed string gravitons and 
momentum $n$.
 These should be given by
\begin{equation}
\epsilon\epsilon(\phi,\dots,\phi,\overbrace{1,\dots,1}^{k})
\tr(Z^n) 
\end{equation}
and anytime we want to add new gravitons we need to add a trace.

We can build multi-open string states with multi-closed string states.
The closed string states will be added by multiplying the above 
operator by traces of other fields.
Perturbatively this is the picture of a weakly coupled D-brane in 
string theory. If we ignore back-reaction, the open strings states and 
closed string states do not mix at zero coupling.

This is not all. Even though we have checked the symmetry of 
particle exchange on the D-brane,  we still
 need to show that the states with one quanta 
are linearly independent (and orthogonal) from states with two quanta, 
etc, and also orthogonal with states that have gravitons in them.
This is how the perturbative string states will appear if 
the D-brane is weakly coupled to the background.
 We can only 
expect this statement to be true in the semi-classical limit, so it 
should be only true at leading order in $1/N$. The $1/N$ corrections 
will give rise to mixing between these states, already at the free field 
theory level\footnote{The Planck constant is $1/N$ in the dimensional
reduction to five dimensions, 
regardless of the value of the dilaton. Thus the $1/N$ mixing 
represents
gravitations corrections to the D-brane}. 
We need to check this fact.
Also, we need to show that these states saturate all possibilities,
namely that there are no more states close to the configuration
with the same quantum numbers 
that can appear on the field theory side.

These two facts together would provide the statement that the mapping 
of states between the gravitational dual and the SYM theory is a unitary 
transformation at leading order.

For example, we could consider the operator
\begin{equation}
\tilde O =
\epsilon\epsilon(\phi,\dots,\phi, \phi Z^n,\overbrace{1,\dots, 1}^k)
\end{equation}
which has the same quantum numbers as the operator ${\cal O}^{N-k}
{a^\dagger_n D_3}$ and ${\cal O}^{N-k}_{D_3}\tr(Z^n)$. 
Naively all of these three operators have different gauge 
structures in how their indices are contracted, and one might believe
that there are more states on the D3-brane that are apparent from
the supergravity calculation. 

However, there is a clever identity that shows that the last operator
$\tilde O$
is a linear combination of the other two.
To show this, we need to 
consider the tensor structure of 
\begin{equation}
\epsilon_{\mu_1,\dots,\mu_{N-k},\rho_1,\dots, \rho_k}
\phi^{\mu_1}_{\sigma_1}\dots\phi^{\mu_{N-k}}_{\sigma_{N-k}}
\end{equation}
Since $\epsilon$ is completely antisymmetric in it's indices, it follows 
that the tensor above is completely antisymmetric in the 
$\sigma$ indices. Therefore, this operator is the same as
\begin{equation}
\epsilon_{\mu_1,\dots,\mu_{N-k},\rho_1,\dots, \rho_k}
\phi^{\mu_1}_{\tilde \sigma_1}\dots\phi^{\mu_{N-k}}_{\tilde\sigma_{N-k}}
\frac 1{(N-k)!}\delta^{[\tilde\sigma]}_{[\sigma]}
\end{equation}
where $[\sigma]$ is a multi-index. The factor of $(N-k)!$ is put in 
to get the 
norm correctly. Now we remember that 
$\delta^{[\tilde\sigma]}_{[\sigma]}$ can be written in terms of $\epsilon$
symbols, so that we get an expression which is 
proportional to
\begin{equation}
\epsilon\epsilon(\phi,\dots,\phi)_{k;k}
\epsilon\epsilon(1,\dots,1,Z^n)_{k;k}
\end{equation}
The four $\epsilon$ symbols have $N-k$ of their indices contracted with
the matrices $\phi,Z^n,1$, and the other $k$ indices are 
contracted among
the $\epsilon$ symbols directly.
Now, in the second pair of $\epsilon\epsilon$ tensors we have $N-k-1$ 
indices contracted with the identity, so we get that 
\begin{equation}
\epsilon\epsilon(1,\dots,1,Z^n)_{k;k}\sim \delta_{k+1;k+1} (Z^n)
\end{equation}
with one of each of the upper and lower indices contracted with $Z^n$.
Now, if we use the permutation representation for $\delta_{k+1,k+1}$
we get terms where the upper tensor index of $Z^n$ is contracted with
it's lower tensor, and some others where both tensor indices  end
contracted with free indices. 
The first type of term gives us $\tr(Z^n)\delta_{k;k}$, so when we 
contract all of the remaining indices with the other tensor 
we get an operator of the form
$$\epsilon\epsilon(\phi,\dots,\phi,\overbrace{1,\dots,1}^k)$$
while the second type of term, once the indices are
contracted gives us an operator of the form
$$\epsilon\epsilon(\phi,\dots,\phi,\overbrace{1,\dots,1}^{k-1},Z^n)$$
This means that the operator we were looking at is indeed a linear 
combination of the other operators we were considering. This argument 
can be generalized for other states with 
open strings.

Now that we have shown that there are non-trivial 
identifications between 
states we need to check the linear independence and orthogonality
of states that we described to ensure that the two spectra of states 
match, between supergravity with a D-brane and the dual field theory 
calculation.

To check this statement we need to calculate the norm of each of the 
states above in the large $N$ limit, 
where $m,n$ are fixed and finite or
at most of order $\sqrt N$ \footnote{This is the limit  where planarity 
starts to fail, and it is also the limit required for understanding 
the plane wave geometry at finite string coupling 
\cite{KPSS,BN,seven}}

Now, we will show the approximate orthogonality of 
states we have described in the large $N$ limit.
 The strategy we will use to prove this statement is 
done in the following steps: first we contract the $\phi$ and
their dual field $\bar \phi$, giving us a contraction of four 
$\epsilon$ symbols. We will leave the contractions of the fields $Z$
implicit in the following way: we will write the contractions as a vev 
on a Gaussian matrix model.
 Then we will use the identity \ref{eq:sum}
to write the norm of the state 
as a vev of a gauge invariant operator in a Gaussian matrix model. 
We will then estimate this vev in the large $N$ limit, 
and from the estimate the approximate orthogonality of 
states will follow.

Consider for example
\begin{equation}
{\cal O}^{N-k}_{a_m^\dagger a_n^\dagger D3} = 
\epsilon\epsilon(\phi,\dots,\phi,\overbrace{1,\dots,1}^{k-2},Z^m,Z^n)
\end{equation}
To calculate the norm we will do the following two point function
\begin{eqnarray}
{\cal O}^{N-k}_{a_m^\dagger a_n^\dagger D3}(x) {\cal O}^{\dagger,N-k}
_{a_m^\dagger a_n^\dagger D3}(y) |x-y|^{N-k+n+m} \\
= (N-k)!^3
\delta_{k;k} \delta_{\bar k, \bar k}(
\overbrace{1,\dots,1}^{k-2},Z^m,Z^n;\overbrace{1,\dots,1}^{k-2},
\bar Z^m,\bar Z^n)_{Z contracted}\\
= <{\cal O}^{N-k}_{a_m^\dagger a_n^\dagger D3}|{\cal O}^{\dagger,N-k}
_{a_m^\dagger a_n^\dagger D3}>
\end{eqnarray}
The $Z$ can be contracted by thinking of them as matrices in
a Gaussian matrix model with measure $\int [dZ][d\bar Z]
\exp-\tr(Z\bar Z)$. 
Now let us use the expansion \ref{eq:sum} at the terms with maximum 
number of traces \footnote{This is the place were keeping $k$ 
finite in the large $N$ limit simplifies the argument},  so that 
\begin{eqnarray}
\delta_{k;k} \delta_{\bar k, \bar k}(
\overbrace{1,\dots,1}^{k-2},Z^m,Z^n;\overbrace{1,\dots,1}^{k-2},
\bar Z^m,\bar Z^n) =\hskip 5 cm
 \\
(k-2)(k-3)(k-2)! \tr(1)^{k-4} \tr(Z^m)\tr(Z^n)\tr(\bar Z^m)\tr(\bar Z^n)
\nonumber \\
+(k-2)(k-2)! \tr(1)^{k-3}\tr(Z^m\bar Z^m)\tr(Z^n)\tr(\bar Z^n)\nonumber\\
+(k-2)(k-2)!\tr(1)^{k-3}\tr(Z^n\bar Z^m)\tr(Z^m)\tr(\bar Z^n)\nonumber\\
+(k-2)(k-2)!\tr(1)^{k-3}\tr(Z^m\bar Z^n)\tr(Z^n)\tr(\bar Z^m)\nonumber\\
+(k-2)(k-2)!\tr(1)^{k-3}\tr(Z^n\bar Z^n)\tr(Z^m)\tr(\bar Z^m)\nonumber\\
+(k-2)!\tr(1)^{k-2}\tr(Z^n\bar Z^n)\tr(Z^m\bar Z^m)
+(k-2)!\tr(1)^{k-2}\tr(Z^n\bar Z^m)\tr(Z^m\bar Z^n)\nonumber
\end{eqnarray}
In the above we have counted the multiplicity for each term that 
appears in the sum.
Now we can evaluate these terms in the large $N$ limit. To do this, it
is convenient to rescale $Z = \sqrt N z$, and extract a factor of 
$N$ from each trace. We are left with

\begin{eqnarray}\label{eq:norm2}
\delta_{k;k} \delta_{\bar k, \bar k}(
\overbrace{1,\dots,1}^{k-2},Z^m,Z^n;\overbrace{1,\dots,1}^{k-2},
\bar Z^m,\bar Z^n) =\hskip 5 cm\\
 N^{k+n+m}(k-2)! <\frac 1 N\tr(z^m\bar z^m)
\frac 1N\tr(z^n\bar z^n)>+\dots
\end{eqnarray}
now, in the Gaussian matrix model in the large $N$ limit we can use 
factorization of correlation functions 
to estimate the above correlator. It turns out that the only 
contribution is exactly from the term left in the line above (we assume 
$n\neq m$), which is
of order $1$ ( a function of $n,m$ alone, with a power series in 
$N^{-2}$).
Once the leading power of $N$ is extracted, the 
terms encompassed by the $\dots$ symbol have zero vev in the strict 
large $N$ limit (fewer traces have suppressions of $1/N$),
 so they are subleading with respect to the first 
term. The $N$ counting of the subleading terms 
is straightforward: each trace counts as a disconnected disk. 
Contractions between different traces bring factors of $1/N^2$
from handles joining the disconnected disks.

The terms with fewer traces also contribute to the norm. For each 
trace we get a factor of $N$, so terms with fewer traces will have 
a $1/N$ suppression.
From the above considerations, the subleading 
contribution to the norm will come from the correlator with 
the following trace structure 
$\tr(1)^{k-2}\tr(z^m\bar z^n z^n\bar z^m)$. In general we will obtain a 
power series in $N^{-1}$ and not $N^{-2}$.
This suggests already that if there is an interpretation in terms
of Riemann surfaces suppressed by the genus, there are contributions 
of worldsheets with boundaries.

Now let us consider an overlap between the state above, and another 
state, also with two quanta but with different momenta 
$\tilde n, \tilde m$, but with the same momentum
$\tilde n+\tilde m = n+m$. 

We want to consider the overlap in the normalization of the states
\begin{equation}
<{\cal O}^{N-k}_{a_m^\dagger a_n^\dagger D3}|{\cal O}^{\dagger,N-k}
_{a_{\tilde m}^\dagger a_{\tilde n}^\dagger D3}>
\end{equation}
We repeat the steps outlined above, and remember to keep the terms 
with maximum number of traces. 
We will thus get contributions of the general
form
\begin{equation}\label{eq:norm1}
N^{k+m+n}< \frac 1N\tr( z^m\bar z^{\tilde m}) 
\frac 1N\tr(z^n z^{\tilde n})> \sim N^{k+m+n-2}
\end{equation}
because at leading order in the large $N$ limit the one point functions
$<\frac 1N\tr(z^n z^{\tilde n})>$ vanish. Thus they start contributing 
at order $N^{-2}$.

However, when we consider terms with fewer traces, there are terms 
of the type
\begin{equation}
\tr(1)^{k-2}\tr(Z^m \bar Z^{\tilde m} Z^n\bar Z^{\tilde n})+
(n\leftrightarrow m)
\end{equation}
which are of order $N^{k+m+n-1}$. It takes a while to see
which contractions  contribute,  but it is found that
the  correlator does receive 
contributions from planar diagrams.

Thus the matrix of overlaps 
for the states
above is roughly of order
\begin{equation}
<O|O^\dagger> \sim N^{k+m+n} 
\begin{pmatrix}
O(1) & O(1/N)\\
O(1/N) & O(1)
\end{pmatrix}
\end{equation}

This shows that the two states are approximately orthogonal. Notice that 
for these operators, the corrections to the norm 
and mixing both 
give an expansion in $1/N$ and not $1/N^2$, consistent with an 
open string interpretation. 

In general it can be shown that the norm of an operator with total
momentum $n = n_1+\dots+n_s$ is of order 
$(N-k)!^3 N^{k+n}$ times factors 
of order one.  If one considers another operator with the same 
momentum, the norm will also be of roughly the same size. 
To consider the overlap we will find that in the maximum number of traces
at least
one of the traces
will contain different powers of $z,\bar z$. Thus the matrix model 
factorized amplitude vanishes, and only the subleading terms 
contribute. These can have less traces, and like the case above we can 
obtain overlap terms which are of order  
$1/N$ with respect to the original norm of the states. 

Now, let us consider comparing a state of a D-brane plus a closed 
string state with one of a D-brane with an open string state.
For example, $a_n^\dagger |D_3>$ and $|D3>\times O_{Closed}$.
These operators are respectively
$\epsilon\epsilon(\phi,\dots,\phi, 1,\dots, 1, Z^n) $
and $\epsilon\epsilon((\phi,\dots,\phi, 1,\dots, 1, 1)\tr(Z^n)$.
The norm squared 
of the first state is of order $(N-k)!^3 N^{n+k}$, which is of 
the same order of magnitude as the second.
The overlap amplitude reduces to the matrix model computation 
of
\begin{equation}
\delta_{k,k}\delta_{\tilde k, \tilde k}
(1,\dots 1, Z^n; 1,\dots 1) \tr(\bar Z^n)
\sim \tr(1)^{k-1} N^2 \tr( z^n) \tr(\bar z^n) \sim N^{k-1} 
\end{equation}
 From here, we see that the overlap of D-branes with an open 
string state and 
a D-brane with a closed string state are also suppressed by $1/N$, and 
hence at leading order the states are orthogonal, but the mixing is again 
of order $1/N$ and not $1/N^2$

In general, for closed string states the contribution from 
factorized amplitudes with the maximum number of traces 
in the matrix model vanishes, due solely to the presence of the 
closed strings. The correction for a state with $s$ gravitons 
is a factor of $N^{-2s}$, but we also have $2s$ more traces to consider
(the traces of the graviton states and their's complex conjugates), 
so the $N$ 
counting for the norm is 
still $N^{k+n}$ where $n$ is the 
total momentum. 

Now, let us consider a state of the form
\begin{equation}\label{eq:osc}
\epsilon\epsilon(\phi,\dots,\phi, \overbrace{1,\dots,1}^
{k}, Z^m\phi Z^n)
\end{equation}
with one $\phi$ sandwiched between two words of the type $Z^k$.
We will ask whether this state is approximately orthogonal to a state
\begin{equation}
\epsilon\epsilon(\phi,\dots,\phi,\phi, \overbrace{1,\dots,1}^
{k-1}, Z^{m+n})
\end{equation} 
or to a state
\begin{equation}
\epsilon\epsilon(\phi,\dots,\phi,\phi, \overbrace{1,\dots,1}^
{k-2}, Z^{m}, Z^{n})
\end{equation} 
These are the two types of $Z$ configurations which can maximize the 
contributions from the ``planar diagrams'' associated to $Z$.

If it is, we can interpret the defect $\phi$ sandwiched between 
the $Z$ as starting to build the open string oscillators 
of the string with 
polarization along $\phi$, and momentum $n+m$
\footnote{In the plane wave  limit this should become
precise when we consider the wave functions of the defect inside the 
string of $Z$'s}.

The first thing we need to take care of is the norm of the state.
We need to worry whether the norm comes primarily from contracting the 
two $\phi,\bar\phi$ which are special, or from ``mixing'' with the 
other $\phi$. It can be easily shown that the contribution without 
mixing (with maximum umber of traces) 
reduces to the Gaussian matrix model vev
\begin{equation}
(N-k)!^3 N^{k+m+n+1}<1/N\tr(z^n x z^m \bar z^m \bar x \bar z^n)>
\end{equation}
with two matrices.
This is essentially the same type of expression for this states
as \ref{eq:norm1}. Here the special $\phi$ has been replaced by 
the matrix model variable $x$.

The mixing term comes as 
\begin{equation}
(N-k-1)!^3 (N-k)^2 N^{m+n+1}
\delta_{k+1,k+1}\delta_{k+1,k+1}(1,\dots 1,\phi, Z^n\phi Z^m;1,\dots 1,\bar\phi, \bar
Z^m\bar\phi Z^m)
\end{equation}
which gives us a set of two terms with maximum number of traces
\begin{eqnarray}
(N-k)!^3 N^{-1}N^{k-1} [\tr(\phi \bar
Z^m\bar\phi \bar Z^n) \tr(\bar\phi Z^n\phi Z^m)\\
+:\tr(\phi\bar\phi):\tr(\bar Z^n\phi Z^m \bar
Z^m\bar\phi \bar Z^n)]
\end{eqnarray}

The normal ordered symbol above is to remind us to contract the 
special $\phi$ with the $\bar\phi$ which are contracted with the 
$\epsilon\epsilon$ tensors.
The first term is of order at most $N^{n+m+2}$ if we use planar 
diagrams, and this is suppressed at $1/N$ with respect to the
calculation above. These planar terms actually come from contracting 
the two exceptional $\phi$'s together, so part of them are just 
subleading because of trace structure. The terms with the mixed 
contractions are suppressed at least by $N^{-2}$ with respect to 
these, which gives a total suppression of order $N^{-3}$.
The second term we need to keep the terms with mixed 
contractions and they give $(N-k)!^3N^{k-2}N^{n+m+2-2}$, so these are 
suppressed by $N^{-2}$.

Orthogonality with respect to the other operators reduces to evaluating
the following
\begin{equation}
\delta_{k,k}\delta_{k,k}(1,\dots,1,Z^n\phi Z^m;
\bar Z^n,\bar Z^m,\bar\phi, 1,\dots 1)
\end{equation}
and 
\begin{equation}
\delta_{k,k}\delta_{k,k}(1,\dots,1,Z^n\phi Z^m;
\bar Z^{n+m},\bar\phi, 1,\dots 1)
\end{equation}
and it is easy to see that these are suppressed 
contributions, either because we have less 
traces, or there are non-planar diagrams to 
evaluate. 
The rest of the calculation is pretty straightforward and 
orthogonality follows.

It is easy to extend this argument to the presence of more $\phi$
which are sandwiched between the $Z$, and also to show that 
configurations with $\phi$ in different positions in the word are 
also orthogonal to each other. In the end, we build string states 
by replacing one of the $1$ symbols with a word, with the additional
condition that it can not begin or end with the letter $\phi$ 
which builds the giant graviton. Because we turn the problem in the end 
to a Gaussian correlation function in the large $N$ limit, different
orders of the letters in the word matter, and different words produce 
different states.

To consider fluctuations orthogonal to the brane along the $AdS$ 
directions, we can attach 
a covariant derivative $(Z^m (D_\mu Z)Z^n)$. More covariant
derivatives in the same word
results in large planar anomalous dimensions, so these build 
oscillators and not
``spherical harmonics'' along the $AdS$ space. This is what shows that 
the D-brane is localized in the AdS directions.
The zero mode $D_\mu$ is
special. We have to contract it with one of the $\phi$, but this 
just generates the descendants of the conformal primary.
We can think of this property as giving the correct boundary 
conditions for the oscillators that are transverse to 
the brane along the $AdS$ directions, showing that we
have only one massless mode along $AdS$ and not infinitely 
many.
Notice that in some sense 
the boundary condition for $\phi$ 
is implemented by the identities we considered, where 
a $\phi$ at the end of the open string
should not be considered as part of the open string itself.

The planar diagrams associated to the words of each string will 
produce the same type of perturbation series for anomalous dimensions
than closed strings do. In this sense, if traces generate a full string 
spectrum with all possible polarizations, it should be the case that 
the planar anomalous dimension for words do the same thing.

This concludes our proof that the combinatorics of the 
spectrum of states of 
large subdeterminants has the right properties to be
describing a single giant graviton which is localized in 
$AdS$ space and that the open 
string states have enough polarizations to
generate the full tower of oscillators.

We need to supplement this with having the geometric
oscillations on the D-brane
to be approximately BPS. This can not be checked in free field theory
nor using just planar diagrams. 
Also, the statements about large anomalous dimensions for 
insertions of various covariant derivatives, or different word orderings
 in a single slot 
are heuristic. A full check is beyond the scope of this paper. 

The assumption we have made above is that if we can neglect the fields
$\phi$ which make the giant graviton to calculate an anomalous
dimension (except perhaps for boundary conditions), then we have 
the requisite number of sates to generate a full open 
string spectrum on the D-brane.
We will now
show that the one loop anomalous dimension of the operators
\ref{eq:onequanta} is protected. This shows that the giant graviton
operator has the correct shape, and it will provide evidence that 
Feynman diagrams involving the giant graviton $\phi$ are suppressed
in the large $N$ limit and lead to a well defined $1/N$ 
expansion.

\section{One loop anomalous dimensions}\label{sec:oneloop}

When considering the large $N$ limit of a theory, it
is important to ask whether perturbation theory for a particular object 
is useful at 
all. Witten has argued that in generic large $N$ theories  baryonic
states will have a mass of order $N$\cite{Wittenbar}, 
which suggests that in general 
the standard perturbation theory will not be well behaved, because 
diagrammatic corrections will grow with powers of $N$.

In the case we want to analyze there is room for improvement 
on this general statement. The main reason is that the state we start 
with is BPS, which in perturbation theory should 
imply that even though the 
individual contributions of a given diagram grow with $N$, 
the contributions to the mass of a state
cancel when all Feynman graphs that contribute 
are added together.

When we consider the operators like \ref{eq:onequanta} and 
\ref{eq:twoquanta} the leading contributions (of order $N$) 
to the anomalous dimensions will cancel because of supersymmetry, 
and then the question is how suppressed are the diagrams that 
contribute at the subleading order to the dimension of the operator.

If they are sufficiently suppressed there are two possibilities:
the result exactly cancels, but it might require small mixing with 
other states and there is a $1/4$ BPS
operator which is near the one that we started, or the perturbative 
expansion is valid even in the strong 't Hooft coupling limit, and 
the expansion is suppressed 
with respect to the natural 't Hooft coupling.
It is very plausible that some of these states are BPS, as multi trace 
$1/4$ BPS operators
which are protected have been found in \cite{Ryzhov, HHHR}.

Now, we turn to the calculation of the anomalous dimension itself.
The first thing we notice is that the operator as described is 
composed only of chiral fields, so there is a chance that the operator
is non-vanishing in the chiral ring. For these chiral operators the 
D-term and vector exchange 
contributions cancel each other\cite{FMMR}, 
and only the F-terms are necessary
to calculate the anomalous dimension of the operator.
When we do a first order perturbation theory calculation, we need to 
choose one of the $\phi$ from the giant graviton
operator to contract with the 
perturbation. Once we find the logarithmic contribution to the two point function, it needs to be contrasted with the norm of the 
state to find the anomalous dimension.

To calculate the norm, we can also choose to write 
the norm in the following form
\begin{equation}
|O^{N-k}_{a_n^\dagger|D_3>}|^2 \simeq \frac{(N-k)!^3 }{(N-k)^2}
\delta_{k+1,k+1}\delta_{k+1,k+1}(\phi, 1,\dots,1 ,Z^n;\bar\phi, 1\dots 1,
\bar Z^n)_{mm} 
\end{equation}
where the right hand side is a Gaussian matrix model calculation.
The $Z$ and $\phi$ are all contracted.
 The main contribution to the norm
comes from the maximum number of traces, and it is
given by
\begin{eqnarray}
|O^{N-k}_{a_n^\dagger|D_3>}|^2\simeq\frac{(N-k)!^3 }{N-k}(k-1)!
N^{k-1}\tr(\phi\bar\phi)\tr(Z^n\bar Z^n)\\
\sim \frac{(N-k)!^3 }{(N-k)^2} (k-1)!N^{k-1+(1+n)+(1+1)}
\end{eqnarray}
and to leading order in $N$ it can be seen that his calculation agrees 
with \ref{eq:norm2}.

To calculate the one loop anomalous dimension, we can write all the $x$
dependence in a factorized form and we can reduce the problem to 
a matrix model computation. With the same normalization of the fields as
above, the anomalous dimension will be proportional to the ratio
\begin{equation}
\frac{(N-k)<\delta_{k+1,k+1}\delta_{k+1,k+1}(\phi,\dots,Z^n;
\bar\phi,\dots Z^n) g_{YM}^2:[X,\phi][\bar Z,\bar\phi]:>_{mm}}
{<\delta_{k+1,k+1}\delta_{k+1,k+1}(\phi,\dots,Z^n;
\bar\phi,\dots Z^n)>_{mm}}
\end{equation}
the $::$ notation indicates normal ordering, no self contraction of the 
F-term vertices. Most of the combinatorial factors cancel between the 
top and bottom lines, as they involve the same contractions of the 
fields $\phi,\bar\phi$ between the giant gravitons. The only 
non-vanishing combinatorial factor ($N-k$) is the one associated
 to choosing 
one field $\phi$ from the giant graviton to contract
 with the perturbation.

The naive $N$ dependence of the interaction vertex 
above is of order $g_{ym}^2N^3$, 
or $\lambda N^2$. If we take into account the non-planarity of the 
contractions we get a $1/N^4$ suppression, so the total contribution to 
the anomalous dimension is of order $g_{YM}^2N^4/N^4\sim g_{YM}^2$.
We need to consider the top contribution in more detail.
The trace structure of the matrix model computation is as follows
\begin{eqnarray}
[\tr(1)^{k-1}[\tr(Z^n\bar Z^n)\tr(\phi\bar\phi)
+\tr(Z^n\bar\phi)\tr(\phi\bar Z^n)\\
-\tr(Z^n\bar Z^n\phi\bar\phi)-\tr(Z^n\bar\phi\phi\bar Z^n)]
+\dots]:\tr([Z,\phi][\bar Z,\bar\phi]):
\end{eqnarray}
the terms denoted with $\dots$ will have less powers of $\tr(1)$,
so they will be suppressed, and can have more traces, but then the 
non-planar suppression is even higher.

One can check that even though the terms of the form $\tr(Z^n\bar\phi)$
only contributes at a subleading order to the norm of the state, when one
considers the contribution to the anomalous dimension of the 
operator it actually cancels the contribution from the first term, 
because contractions with the interaction vertex 
come with opposite signs. Also, 
the terms which could give a bigger contribution to the anomalous 
dimension than the first two (because of this cancellation) are 
doubly non-planar 
suppressed as well, 
because in the vertex $:\tr([Z,\phi][\bar Z,\bar\phi]):$ the
two chiral terms appear together, whereas in the trace structure of the
matrix model trace 
they have an order where the chiral and anti-chiral fields are 
alternating. 

Putting all of the calculation together we see that the anomalous 
dimension of the operator is of order at most $g^2N^{-1}$ to leading 
order. This argument 
can be extended to having more open strings with their momentum along
$Z$. The result is that at least to leading order, perturbation theory 
for these states gives very small corrections, and they become smaller 
in the large $N$ limit. We take this as evidence that the identification 
of states we have done can interpolate between strong and weak 't Hooft 
coupling. Clearly, this matches the result from fluctuations 
around the giant graviton \cite{Jetal}

\section{Organizing the $1/N$ expansion.}
\label{sec:org}

In this section we will give a description of how to organize 
the $1/N$ expansion of the giant graviton operators with their 
stringy fluctuations in general.

The generic operator we have been discussing 
can be written in the following form
\begin{equation}
O^{N-k} \sim \epsilon\epsilon(\phi,\dots,\phi, 
\overbrace{1,\dots,1}^{k-l},w_1,w_2,\dots, w_l)\tr(s_1)\dots \tr(s_m)
\end{equation}
where the $w_l, s_i$ are words, and $w$ does not begin or end in 
the symbol $\phi$. 

Let us consider first the case where there are no fields $\phi,\bar\phi$ 
in the 
words $w_i, s_l$. Then the free contraction of two of these operators 
(after we contract all the $\phi,\bar\phi$)
will give rise to the following Gaussian matrix model amplitude 
\begin{equation}\label{eq:expr1}
(N-k)!^3<\delta_{k;k}\delta_{\bar k;\bar k}
(1,\dots,1, w_1, w_2\dots w_l;1,\dots 1,\tilde w_1, \dots \tilde , w_{a})
\tr(s_1)\dots \tr(s_m) \tr(\tilde s_1)\dots \tr(\tilde s_b)>
\end{equation}

This expression above is some sum of traces where each 
word in the expression has a finite number of letters (or of order 
up to $\sqrt(N)$ if one wants to study the plane wave limit).
This expression \ref{eq:expr1} has a well defined $1/N$ expansion.
The individual trace combinations have a $1/N^2$ expansion, but since 
the expression above has various trace types, these give rise to a
$1/N$ expansion.

Similarly, we can consider expressions 
where some of the words have fields
$\phi$, $\bar\phi$ (but not at the beginning or end of the $w_i$), and 
it can also be the case that $k\neq \tilde k$, but 
that $k-\tilde k$ is finite. From here, most of the $\phi$, 
$\bar \phi$ will be contracted as usual, but let us assume that there are
$t_1, t_2= t_1+k-\tilde k$ of these which are 
contracted with the words inside the 
traces.

These terms will give rise to the following type of matrix model 
amplitudes
\begin{eqnarray}
(N-k-t_1)!^2 C(k,t_1,t_2)
\delta_{k+t_1,k+t_1}
\delta(\tilde k+t_2,\tilde k+t_2)
(\overbrace{\hat \phi,\dots,\hat\phi}^{t_1},1,
\dots,1,w_1,\dots, w_n; \nonumber\\
\overbrace{\hat{\bar \phi},\dots,\hat{\bar\phi}}^{t_2}, 1,\dots
\tilde w_1,\dots \tilde w_a)\tr(s_1)\dots\tr(s_l)
\tr(\tilde s_1)\dots \tr(\tilde s_b)
\end{eqnarray}
where the $\hat \phi$ fields are not allowed to be contracted with 
the $\hat{\bar\phi}$ fields. In this notation, $C(k,t_1,t_2)$ 
is a combinatorial 
constant that counts the number of contractions of this type.

This constant is given by 
\begin{equation}
C(k,t_1,t_2)= \frac{(N-\tilde k)!}{t_2!}
{N-k \choose t_1}
\end{equation}
It can be checked explicitly (using $N-k-t-1 = N-\tilde k-t_2$) that the 
above expression for $C$ is symmetric under the exchange of the 
two operators one can be analyzing.

Now, we want to show that as we sum over all possible values of $t_1$, 
successive terms give rise to terms of order  $1/N^{t_1}$ 
in the $1/N$ expansion.

The total combinatorial coefficient in front of the operator is 
\begin{equation}
\frac{(N-\tilde k)!(N-k)!(N-k-t_1)!}
{t_2!t_1!} 
\end{equation}
which when compared to the same expression with different
values of $t_1$ is of order 
\begin{equation}
(N-k-t_1)!/(N-k-\tilde t_1)!
\sim \frac{(N-k)!/(N-k)\dots (N-k-t_1+1)}
{(N-k)!/(n-k)\dots (N-k-\tilde t_1+1)}
\sim N^{- (t_1-\tilde t_1)}
\end{equation}
with the coefficient having a well defined $1/N$ 
expansion.

Now, we need to understand the contractions of the $\hat \phi$ and the 
$\hat{\bar \phi}$. First of all, there are configurations with
more traces which are possible, $k+t_1$ of them. This gives a factor of 
$N^{k+t_1}$ from counting traces. Also, each pair $\phi$, $\hat{\bar\phi}$
gives an extra factor of $N$, so combining all 
the factors of $N$ it seems 
as if these amplitudes are of order $N^{2t_1-t_1}$, times some fixed 
power of $N$. This is the combinatorial growth of Feynman diagrams that 
is responsible for possible bad behavior in the $1/N$ expansion.

However, the definition of the operators should be in normal 
ordered form if selfcontractions of the operator are allowed (this is 
automatic for chiral operators), and we have also set the rule that
the $\hat\phi$ can not be contracted with the $\hat{\bar\phi}$, 
so that the $\hat\phi$ can only 
contract with letters in the $\tilde w$ and $\tilde s$, and similarly 
the $\hat \phi$ only contract with the $w$ and $s$. 
 Also, $w$ is such 
that it can not begin or end on a $\phi$ operator, so if the
$\hat{\bar\phi}$ appears in the same trace as 
a $w$ all of the possible contractions are non-planar in the matrix model.
Because of these non-planarity, one gets at least an extra factor of 
$N^{-2t_1}$ with respect to planar amplitudes, and the simplest case is
to consider contractions where the words $w,\tilde w$ are missing.

All together, the amplitude is suppressed in $N^{-t_1}$.
Notice also that when we include interactions, the interaction vertex
is of order $N^2\lambda$, and can contribute to self-energy diagrams
at order $\lambda$ (this is because the contraction will be non-planar
with respect to the large $N$ vev of the 
perturbation), which have the same combinatorial structure as 
above, 
but without the additional 
non-planar suppression. However, supersymmetry should 
ensure that the pure self energy
of the $\hat\phi, \bar\hat \phi$ will cancel (because these diagrams
 would contribute to the anomalous dimension of the protected half-BPS
 operator otherwise), so diagrams with self-energy  
contributions are also suppressed by powers of $N^{-t_1}$, because they 
can only 
involve dressed propagators for the same diagrams as we had above.

Also, if we study the trace structure of operators of the type above 
in the presence of interactions, the interactions leave most of the 
fields $\phi$ untouched, and if we think of them as partons, their
color structure is still completely antisymmetric in lower and 
upper indices for all but a finite number of the $\phi$. This means that
in perturbation theory if we begin in the class of operators we have 
been studying, we stay in this class of operators. In this sense, 
perturbation theory is closed within the class of operators we have 
been studying.

Mixing with operators of different trace structure (combinations
of giant gravitons with distinct sizes for example) are allowed. 
However, if they are not of the shape we have been describing, their 
overlap with the given operators above is suppressed much more than any
polynomial behavior in $N$. For other combinations of giant 
gravitons it is exponentially suppressed in $N$, and mixing 
with these states is to be considered as non-perturbative.
Thus, in perturbation theory we have described a sector which can 
be studied independently and has a well defined $1/N$ expansion.

Putting all of the information together: the perturbation theory 
around the giant graviton operators seems to be well defined, 
and it is described by a sector of the large $N$ limit of the 
${\cal N}=4$ SYM. Clearly, the argument above is sketchy as we have 
not discussed issues with mixed combinatorics, 
``overlapping divergences''. To first order it is correct.
It is certainly worthwhile to explore this expansion in more detail, 
but it is beyond the scope of the present article.

To summarize, there are two main effects contributing 
for the large $N$ counting: combinatorial 
enhancement of diagrams, together with non-planar suppressions.
The planar diagrams contributing to the self-energy of the 
``background'' 
D-brane 
itself cancels, and only diagrams that mix the D-brane and the 
defects on it contribute. The non-planar suppressions induced by these 
diagrams dominate over the combinatorial enhancement, and the 
D-brane with defects has a well defined $1/N$ expansion.

Considering also the combinatorial structure of planar anomalous 
dimensions together with restrictions on the shape of operators, 
the D-brane is producing 'boundary conditions' for letters in the
words $w_i$. These planar diagrams are of the same type as for closed 
string states, and thus should lead to the same type of spectrum 
of anomalous dimensions (strings with boundary conditions).

\section{Conclusion}

In this paper we have shown that the conjectured dual operator to a
giant graviton has all the requisite properties to describe a D-brane.
This is, there is a collection of operators which can be considered 
as stringy excitations of the D-brane \footnote{The
spectrum of fluctuations of the D-brane state 
approximates a Fock space of string states in the large $N$ limit}, and 
it has the correct spectrum of low energy (massless) excitations on it's 
worldvolume. This constitutes a test of shape of the D-brane. 
This is important as it helps us understand the nature of locality in
higher dimensions from the holographic dual theory.
 
Also, we have presented evidence that the perturbative diagrams 
contributing to stringy amplitudes have a well 
defined $1/N$ expansion, which can be made explicit 
by using matrix model techniques.

Clearly these combinatorial properties will be the same for other SCFT
in four dimensions, except that some of the exact details will have to 
change because of the different matter content of the theory.

Supersymmetry, as in other examples \cite{BMN}, 
was the key to understanding the giant graviton 
operators: without self energy diagram cancellations 
the perturbation 
theory does not seem to behave well at all, and would conform to 
a more standard picture of baryons in large $N$ theories 
\cite{Wittenbar}.

Many puzzles still remain open.

It would be 
interesting if one could show that there is a spectrum of excitations 
which is $1/4$ BPS protected. This might require taking into account 
some amount of mixing between states before one can decide weather this 
is true or not, and judging from \cite{Ryzhov,HHHR}, this will 
almost certainly 
be the case.

Here we have shown that to leading order there is such a spectrum of 
states. However, it does not determine the position of the D-brane 
fully, as giant gravitons of different sizes have the same spectrum 
of fluctuations independent of their size. It could be the case that 
interactions between the waves can tell them apart.
With the methods presented here 
it should be possible to check the non-linear 
terms in the DBI action. This is, the massless 
open string scattering on the D-brane worldvolume. 

There are
other configurations of giant graviton states 
with less supersymmetry in 
the geometric limit \cite{Mikhailov}, which have also been studied
in \cite{Beasley,Ouyang}. If these states will be protected, 
it is probably true that the spectrum of $1/4$ BPS states 
described above should exist.
There is to my knowledge
no candidate dual operators to these other giant graviton states 
in the literature.
Considering that for these type of objects non-planar diagrams dominate, 
it will require a lot of combinatorial dexterity to show that any such 
operator can have a well defined $1/N$ expansion. Even for a giant 
graviton which is less than near-maximal, the combinatorial problem is 
much harder than what has been presented here.

In this paper, we have mentioned 
the possibility of studying the plane wave limit 
in the presence of a giant graviton. This is a natural scenario to 
study as 
it is a limit which simplifies the diagramatics considerably: 
it is known that the spectrum 
of states can be controlled very well, and it would be interesting if
one can compare amplitudes in the presence of the D-brane. These tests 
should finish determining the position of the brane and showing that 
the boundary conditions set by the D-brane operator are the correct ones.
 The reason this is tests the location of the D-brane is because 
the plane wave states are very localized in a small tube inside 
the AdS spacetime, so interactions with the closed string states on the 
plane wave are unsuppressed only if the D-brane contains the null 
trajectory along which one is taking the Penrose limit. 

Finally, it would be very interesting if one could extend these ideas to 
the presence of many D-branes on top of each other. This should serve as a
test on how the $U(N)$ gauge theory on the D-brane worldvolume 
arises. The 
comparison between states in the geometric limit and states in the 
gauge field theory should be such that one obtains the $U(N)$ 
invariant sector of a Fock space of strings in the adjoint of $U(N)$.
Understading the map of states could be a quite complicated 
combinatorial problem. This can be made explicit 
if one considers a simpler case with 
just two D-branes on top of each other,
both with maximum angular momentum and given by the operator 
$O_{D3}^2$. Then the connected diagram 
contribution to the norm of the state dominates over the product 
of the norm of the individual states. In essence, one can not 
consider the state as factorized in any sense, unlike 
when one takes the product of two ordinary graviton states.
It is possible that this behavior is responsible for producing the 
enhanced gauge symmetry when the two D-branes are on top of each 
other, but at this point, this is just 
speculation.

\section*{Acknowledgments}
I would like to thank V. Balasubramanian, S. Cherkis, 
R. Gopakumar, A. Hashimoto, 
I. Klebanov and J. Maldacena for various useful discussions. Research 
supported in part by DOE grant DE-FG02-90ER40542

\appendix
\section{Combinatorial identities}

Here we will derive some combinatorial identities which are used 
in the paper. These are built out of the totally antisymmetric tensor
$\epsilon_{\mu_1,\dots, \mu_N}$ in $N$ entries, which is a gauge
invariant tensor for $SU(N)$.

The components of $\epsilon$ are
\begin{equation}
\epsilon_{\mu_1,\dots, \mu_N}=\left\{\begin{matrix}
\pm 1, \hbox{If all of the $\mu_i$ are different}\\
0 \hbox {Otherwise}
\end{matrix}\right.
\end{equation}
The sign of the symbol is determined as follows
\begin{equation}\label{eq:defeps}
\epsilon_{\sigma(1),\dots, \sigma(N)}= (-1)^{|\sigma|}
\end{equation}
for $\sigma$ a permutation of $1,\dots, N$, and $|\sigma|$ is the 
parity of the permutation.

We define a similar object with raised indices.
 We can similarly define an 
$\epsilon$
tensor with all indices raised.

Many of the calculations in the paper involve the contraction of 
$\epsilon$ tensors. First, consider the contraction of $N-k$ indices of 
two $\epsilon$ tensors
\begin{equation}
\epsilon^{\mu_1,\dots,\mu_{N-k}, \rho_1, \dots, \rho_k}
\epsilon_{\mu_1\dots,\mu_{N-k}, \nu_1,\dots, \nu_k} = 
\epsilon\epsilon^ {\rho_1, \dots, \rho_k}_{\nu_1,\dots, \nu_k}
\end{equation}
The contraction above defined the $\epsilon\epsilon$ symbol.
Now let us evaluate the components of $\epsilon\epsilon$.
We notice immediately that the tensor is completely antisymmetric in
both it's lower and upper indices. So if any of the upper or lower 
indices is repeated the result is zero.

Secondly, once we have fixed $\rho_1,\dots,\rho_k$ all different, 
then for the terms to contribute in the sum we have to sum over 
all possible combinations of $\mu_i$ which are different from the 
$\rho_j$. 
It also follows that the result is zero unless 
the set of the $\nu_j$ is the same set as the $\rho_i$, 
with permutations allowed.

Given such a permutation $\sigma(\rho_j)$, 
we get from \ref{eq:defeps} that when we evaluate the tensor we 
find
\begin{equation}
\epsilon\epsilon^{\rho_1,\dots,\rho_k}
_{\sigma(\rho_1),\dots,\sigma(\rho_k)}= (N-k)!(-1)^{|\sigma|} 
\end{equation}
The $(N-k)!$ is the number of possible different combinations of 
$\mu_1,\dots\mu_{N-k}$ that can contribute. They all contribute with 
the same sign.
Since the tensor is different from zero only when the upper and lower 
indices coincide, and the components are a fixed number (up to a sign), 
we define the standard symbol
\begin{equation}
\epsilon\epsilon^{\rho_1,\dots,\rho_k}
_{\nu_1,\dots,\nu_k}= (N-k)!\delta^{[\mu_1,\dots,\mu_k]}
_{[\rho_1,\dots\rho_k]}
\end{equation} 
and it can also be shown by comparing the components of tensors that 
\begin{equation}\label{eq:iden1}
\delta^{[\mu_1,\dots,\mu_k]}
_{[\rho_1,\dots\rho_k]} =\sum_{\sigma}
(-1)^{\sigma} 
\delta^{\mu_1}_{\rho_{\sigma(1)}}\dots\delta^{\mu_k}_{\rho_{\sigma(k)}}
\end{equation}
We will also use the notation $\epsilon\epsilon_{k,k}$ and 
$\delta_{k,k}$ as a multi-linear operator with $k$ free upper 
indices and $k$ free lower indices. If $k=0$, the symbol 
is a pure number $\epsilon\epsilon_{0,0}= N!$.

We have the following identity if we introduce matrices
\begin{equation}\label{eq:multil}
\epsilon\epsilon_{N,N}(1,\dots, 1, M_1,\dots, M_k)
=\epsilon\epsilon_{k,k} (M_1,\dots, M_k)
\end{equation}
We can also write the above in terms of the $\delta_{k,k}$ symbol
\begin{eqnarray}
\delta_{k,k}(M_1,\dots, M_k)
&=&\delta^{[\rho_1,\dots,\rho_k]}_{[\nu_1\dots\nu_k]}
(M_1)_{\rho_1}^{\nu_1}\dots (M_k)_{\rho_k}^{\nu_k}\\
&=&\sum_{\sigma}
(-1)^{\sigma} 
\delta^{\mu_1}_{\rho_{\sigma(1)}}\dots\delta^{\mu_k}_{\rho_{\sigma(k)}}
(M_1)_{\rho_1}^{\nu_1}\dots (M_k)_{\rho_k}^{\nu_k}\label{eq:paren}
\end{eqnarray}
Now, if we think of $\sigma$ as a permutation of the set $\{1,\dots,k\}$, 
we can write the permutation in terms of a cycle representation 
$\hat\sigma = 
(i\sigma(i)\sigma^2(i)\dots)(j\sigma(j)\sigma^2(j)\dots)\dots$
where $i\to\sigma(i)\to\sigma^2(i)$ etc. 
The way the indices are contracted in the expression above
\ref{eq:paren} give rise to traces, one for each 
parenthesis in the cycle representation of $\sigma$.
The above then gives 
\begin{equation}\label{eq:paren2}
\delta_{k,k}(M_1,\dots, M_k)=
\sum_\sigma (-1)^{|\sigma|}
\tr(M_iM_{\sigma(i)}M_{\sigma^2(i)}\dots)\tr(
M_jM_{\sigma(j)}\dots)\dots
\end{equation}
an alternating sum over all possible trace combinations of 
the matrices $M_1,\dots, M_k$, determined by the permutations of 
$\{1,\dots, k\}$.

In particular, if for each trace we associate a factor of $N$, then the 
above expression is a $1/N$ expansion in terms of sums of traces, 
starting with
\begin{equation}
\delta_{k,k}(M_1,\dots, M_k) = N^k\left(\frac{\tr(M_1)}{N}\dots 
\frac{\tr(M_k)}{N}\right)
-N^{k-1}\sum_{i<j}\left(\frac{\tr(M_iM_j)}{N}\prod_{s\neq i,j}
\frac{\tr(M_s)}{N}\right)+ \dots
\end{equation}

Now, we will consider the combinatorics of contracting
four $\epsilon$ symbols. 
Let us consider 
$\epsilon\epsilon_{k,k}\epsilon\epsilon_{k,k}$ with all indices 
contracted between the two $\epsilon\epsilon$ symbols. This is equal to
\begin{eqnarray}
\epsilon\epsilon_{k,k}\epsilon\epsilon_{k,k}&=&
\epsilon^{\mu_1,\dots,\mu_N}
\epsilon_{\mu_1,\dots,\mu_{N-k},\sigma_{N-k+1},\dots
\sigma_N}\epsilon^{\sigma_1,\dots,\sigma_N}
\epsilon_{\sigma_1,\dots,\sigma_{N-k},\mu_{N-k+1},\dots
\mu_N}\\
&=& (N-k)!^2 \delta_{k,k}\delta_{k,k}\\
&=& N! (N-k)!k!
\end{eqnarray}
The last line 
follows from the fact 
that for the terms to contribute we need that the two 
sets
$\{\mu_1,\dots,\mu_{N-k}\}$ and $\{\sigma_1,\dots,\sigma_{N-k}\}$ are 
identical, and that all terms contribute with the same 
sign.
The number of ways of choosing indices for $\mu_1,\dots,\mu_N$
is $N!$. Now, the indices for $\sigma_1,\dots, \sigma_{N-k}$ 
are fixed up to permutations, as well as 
$\sigma_{N-k-1},\dots,\sigma_{N-k}$. 
Each of these gives us an extra multiplicity of $k!(N-k)!$ from 
permutations of the fixed sets. Notice the symmetry $k\to (N-k)$
of the final answer which is obvious from the index contractions.  
From the last line, we see that
\begin{equation}
\delta_{k,k}\delta_{k,k} = k!^2{N \choose k}
=k!N^k(1-1/N)\dots (1-k/N)
\end{equation}
Similar to \ref{eq:multil}, we can consider the $\delta\delta$ or
$(\epsilon\epsilon)(\epsilon\epsilon)$ tensors as 
multilinear 
operators with various entries. We will be particularly 
interested in contracting
the indices with matrices in such a way that all of the upper indices of
one of the $(\epsilon\epsilon)$ are attached
 with the lower indices of the other $(\epsilon\epsilon)$ group.
 
The notation we will use for this is as follows
\begin{equation}
\delta_{k,k}
\delta_{\bar k\bar k}
(M_1,\dots, M_k;\tilde M_1, \dots \tilde M_k)
= \delta^{[\mu_1,\dots,\mu_k]}_{[\rho_1,\dots\rho_k]}
\delta^{[\bar\mu_1,\dots,\bar\mu_k]}_
{[\bar\rho_1,\dots,\bar\rho_k]}
(M_1)^{\rho_1}_{\bar\mu_1}\dots(M_k)^{\rho_k}_{\bar\mu_k}
(\bar M_1)^{\bar\rho_1}_{\mu_1}\dots(\bar M_k)^{\bar\rho_k}_{\mu_k}
\end{equation}
Now, we can use the identity \ref{eq:iden1} to rewrite the above 
contraction as
\begin{eqnarray}\delta_{k,k}
\delta_{\bar k\bar k}
(M_1,\dots, M_k;\tilde M_1, \dots \tilde M_k)&=&
\sum_{\sigma} \delta_{k,k} 
(-1)^{2|\sigma|} \delta_{k,k}(M_1\bar M_{\sigma(1)} \dots
M_k\bar M_{\sigma(k)})\\
&=&\sum_{\sigma}\delta_{k,k}(M_1\bar M_{\sigma(1)},\dots
,M_k\bar M_{\sigma(k)})
\end{eqnarray}
The fact that we get $(-1)^{2\sigma}$ is that once we contract the upper 
index of $M_i$ with the lower index of $\bar M_{\sigma(i)}$ we need 
to take into account the permutation of the upper indices of 
$\bar M_{\sigma(i)}$ with respect to the lower indices of $M_i$.
This permutation is $\sigma^{-1}$, so we get a factor of 
$((-1)^{|\sigma|})^2$ from this extra reshuffling of indices.
Now, we can use again identity  \ref{eq:paren2} to write the above as 
an alternating sum over all trace combinations 
of the composite matrices 
$M_i\bar M_{\sigma(i)}$, and over all possible 
permutations $\sigma$.

\end{document}